\documentclass[aps, prapplied, twocolumn,nofootinbib,superscript address]{revtex4-2}

\usepackage{graphicx}
\usepackage{dcolumn}
\usepackage{bm}
\usepackage{orcidlink}
\usepackage{physics}
\usepackage{mhchem}
\usepackage{empheq}
\usepackage{siunitx}
\usepackage{natbib}
\usepackage{tcolorbox}
\newtcolorbox{mybox}[1]{fonttitle=\bfseries,title=#1}
\newcommand{\ii}{{\rm i}}

\begin{document}

\title{Fr\"{o}hlich versus Bose-Einstein Condensation in Pumped Bosonic Systems}

\author{Wenhao Xu\orcidlink{0009-0000-1854-4047}}
\email{wx266@cam.ac.uk}
\affiliation{Department of Physics and Living Systems Institute, University of Exeter, Stocker Road, Exeter EX4 4QD, United Kingdom}
\affiliation{Department of Biophysics, Radboud University, Houtlaan 4, 6525
XZ Nijmegen, Netherlands}
\author{Andrey A. Bagrov\orcidlink{0000-0002-5036-5476}}
\email{andrey.bagrov@ru.nl}
\affiliation{Institute for Molecules and Materials, Radboud University, Heijendaalseweg 135, 6525 AJ Nijmegen, Netherlands}
\author{Farhan T.\ Chowdhury\,\orcidlink{0000-0001-8229-2374}}
\email{f.t.chowdhury@exeter.ac.uk}
\affiliation{Department of Physics and Living Systems Institute, University of Exeter, Stocker Road, Exeter EX4 4QD, United Kingdom}
\author{Luke D. Smith\orcidlink{0000-0002-6255-2252}}
\affiliation{Department of Physics and Living Systems Institute, University of Exeter, Stocker Road, Exeter EX4 4QD, United Kingdom}
\author{Daniel R.\ Kattnig\,\orcidlink{0000-0003-4236-2627}}
\affiliation{Department of Physics and Living Systems Institute, University of Exeter, Stocker Road, Exeter EX4 4QD, United Kingdom}
\author{Hilbert J. Kappen\,\orcidlink{0000-0002-5728-3676}}
\affiliation{Donders Center for Neuroscience, Radboud University, Heijendaalseweg 135, 6525 AJ Nijmegen, Netherlands}
\author{Mikhail I. Katsnelson\,\orcidlink{0000-0001-5165-7553}}
\affiliation{Institute for Molecules and Materials, Radboud University, Heijendaalseweg 135, 6525 AJ Nijmegen, Netherlands}

\date{\today}

\begin{abstract}
Magnon-condensation, which emerges in pumped bosonic systems at room temperature, continues to garner great interest for its long-lived coherence. While traditionally formulated in terms of Bose-Einstein condensation, which typically occurs at ultra-low temperatures, it could potentially also be explained by Fr\"{o}hlich-condensation, a hypothesis of Bose-Einstein-like condensation in living systems at ambient temperatures. This prompts general questions relating to fundamental differences between coherence phenomena in open and isolated quantum systems. To that end, we introduce a simple model of bosonic condensation in an open quantum system (OQS) formulation, wherein bosons dissipatively interact with an oscillator (phonon) bath. Our derived equations of motion for expected boson occupations turns out to be similar in form to the rate equations governing Fr\"{o}hlich condensation. Provided that specific system parameters result in correlations that amplify or diminish the condensation effects, we thereby posit that our treatment offers a better description of high-temperature condensation compared to traditional formulations obtained using equilibrium thermodynamics. By comparing our OQS derivation with the original uncorrelated and previous semi-classical rate equations, we furthermore highlight how both classical anti-correlations and quantum correlations alter the bosonic occupation distribution. 

\end{abstract}

\maketitle

Bose-Einstein Condensation (BEC) occurs when a distribution of identical non-interacting bosons is heavily dominated by modes occupying the ground state \cite{einstein1924quantentheorie,bose1924plancks}. The simplest example system is an ensemble of uncoupled quantum harmonic oscillators, the grand partition function for which is well-known, with the occupation numbers of each state $k$ following the Planck distribution: $\expval{\hat{n}_k} = 1/(\exp(\beta(\epsilon_k - \mu)) - 1)$, where $\epsilon_k$ is the energy, $\mu$ is the chemical potential, and $\beta = (k_BT)^{-1}$ is the inverse temperature, $\mu\leq \epsilon_k$ must always hold for non-negative occupations in the ground state. When minimal single-particle energy approaches the chemical potential level, $\epsilon_0\to\mu$, ground state occupation diverges, $\expval{\hat{n}_0}\to\infty$, manifesting BEC \cite{sethna2021statistical}. Excluding situations arising in ultra-cold physics, condensation also arises in open quantum systems (OQS) dissipatively interacting with some environment. Since the latter deleteriously affects conditions required to achieve BEC \cite{wheatley}, there is the need to reconcile the open systems emergence of the phenomenon with its atypically long-lived coherence. 

Here, we consider Bose-Einstein-like condensation in non-equilibrium pumped bosonic systems. We are motivated by seminal experiments on the observation of magnon condensation in yttrium-iron-garnet (YIG) films at room temperature \cite{chumak2009bose, demokritov2006bose, demidov2008observation, demidov2007thermalization, dzyapko2007direct}. The phenomenon is ascribed to a quasi-equilibrium state of a Bose gas, induced by microwave pumping at a power exceeding some threshold. Magnons are accumulated at the lowest energy levels characterized by non-zero in-plane wave vectors under an external magnetic field. Nonlinear interactions, including magnon-phonon interactions and four-magnon scattering, are presumed to be crucial for the required thermalization to quasi-equilibrium steady states, but despite almost two decades of research, broadly compelling theory continues to be lacking \cite{noack2021evolution, borisenko2020direct, salman2017microscopic, bozhko2016supercurrent, ruckriegel2015rayleigh, Dzyapko2014, hick2012thermalization, rezende2009theory}. In this letter, we address this from a novel OQS perspective, by considering magnons that weakly interact with a phonon bath, and further connect to Fr\"ohlich condensation, which, hypothesized to describe non-thermal collective properties in living systems, models nonlinearly interacting sub-units supplied by exogenous energy sources at room temperature \cite{frohlich1968bose,frohlich1968long}. Similarly to magnon condensation, when the external energy pumping exceeds a certain threshold, the modes of harmonic oscillators with a discrete energy spectrum predominantly occupy the ground state. Previous efforts attempted to underpin the theoretical foundation of Fr\"ohlich condensation to classical stochastic processes on quantized energy spectra \cite{preto2017semi}. While Fr\"{o}hlich condensation has remained controversial for a perceived lack of experimental demonstrations in specific living systems \cite{lechelon2022experimental, nardecchia2018out, lundholm2015terahertz}, recent efforts have shown its relevance for non-classical systems involving molecular vibrations \cite{schroer2021probing,zhang2019quantum} and phonons in optomechanics \cite{zheng2021frohlich}.

One has to emphasize, however, that we simplify our model in comparison to what is required to describe more directly the above-mentioned experiments. Broadly, they deal with the longitudinal pumping, that is, the pumping of magnon-pairs \cite{dzyapko2007direct}, rather than pumping of single bosons. Since our aim is to clarify the difference between the condensation in open and isolated quantum systems in general, rather than to explain specific experiments, we constrain our attention to describing magnons modelled as pumped bosons. We thus derive the magnon equation of motion and highlight its surprising similarity to the rate equation describing Fr\"ohlich condensation. 
Both the earlier semi-classical and our OQS derivations result in ``correlated" rate equations, where magnon occupations at different wave numbers are correlated. Using a three-mode toy model, we demonstrate that the condensation effect is controlled by these correlations. Classical anti-correlations between two occupation numbers \cite{preto2017semi} lead to a more condensed distribution in the ground state compared to the uncorrelated one, while quantum correlations between two magnon number operators narrow the occupation differences between energy levels. 

Physical properties of YIG films can be described by a Heisenberg-like quantum ferromagnet model on a cubic lattice \cite{rezende2009theory,akhiezer1968spin}, incorporating exchange and dipole-dipole interactions between localized spins, and the Zeeman interaction with an external magnetic field \cite{bagrov2020multiscale, PhysRevLett.117.137201, prudkovskii2005topological}. The spin Hamiltonian can be converted to magnonic oscillator basis using the Dyson-Maleev or the Holstein-Primakoff transformation, which are equivalent at the second-order expansion and large-$S$ limit \cite{akhiezer1968spin,dyson1956general,dyson1956thermodynamic,dembinski1964dyson}. Further, through Fourier and Bogoliubov transformations, we diagonalize the second-order quadratic Hamiltonian to be quantum harmonic oscillators in the wave-number representation
\begin{align}
    \hat{H}_{\rm S}&= \sum_{\mathbf{k}}\omega_{\mathbf{k}}\hat{c}_{\mathbf{k}}^{\dagger}\hat{c}_{\mathbf{k}},
    \label{eq:free-magnon}
\end{align}
where $\omega_{\bf k}$ is magnon spectral frequency, and $\{\hat{c}_{\mathbf{k}}^{\dagger},\hat{c}_{\mathbf{k}}\}$ are bosonic creation and annihilation operators of magnon oscillators at wave number $\vb{k}$ (see SM for details). We consider second and third-order magnon-phonon interactions that dissipate free magnons and contribute to the condensation. The interactions stem from the fluctuations of exchange couplings due to lattice displacements \cite{bozhko2017bottleneck,kreisel2009microscopic,ruckriegel2014magnetoelastic}, as seen in
\begin{align}
\begin{split}
    \hat{H}_{\rm m,ph}=&\sum_{{\bf k},\sigma}V_{\sigma{\bf k}}\hat{b}_{\bf k\sigma}\hat{c}_{\bf k}^{\dagger}+\text{h.c.}\\
    &+\frac{1}{\sqrt{N}}\sum_{\bf k,\sigma}\sum_{\bf k'}V_{\sigma{\bf kk'}}\hat{c}_{\bf k}^{\dagger}\hat{c}_{\bf k'}\hat{b}_{\bf k-k'\sigma}+\text{h.c.},
\end{split}\label{eq:magnonphononinterac}
\end{align}
where $\sigma=\{\parallel,\perp_1,\perp_2\}$ labels three acoustic phonon polarization, $\omega_{\bf k\sigma}$ are phonon frequencies, $\{\hat{b}_{\bf k\sigma}^{\dagger},\hat{b}_{\bf k\sigma}\}$ are bosonic creation and annihilation
operators of phonon eigenmodes with momentum ${\bf k}$, and $V_{\sigma{\bf k}}$ and $V_{\sigma{\bf kk'}}$ are the corresponding coupling strengths.
\par 
Experimentally, magnon condensation can be demonstrated when systems are pumped by a microwave resonator with large enough power \cite{demokritov2006bose,demidov2008observation,dzyapko2007direct}. In real experiments, longitudinal pumping was used, that is, creation of magnon pairs by external electromagnetic radiation. It does not seem to be apt to consider directly this complicated model, before studying the simple case of single-magnon pumping with amplitude degeneracy ($\abs{g_{\bf k}}=\abs{g_{\bf -k}}$ in Eq. \eqref{eq:classicalpump_main}). With the latter case being so far unexplored, we devote our attention to it in order to provide insight into magnon-phonon interactions. 

Consider a monochromatic microwave, formulated as a quantum harmonic oscillator with $\hat{H}_{\text{R}}=\omega_{\text{p}}\hat{R}^{\dagger}\hat{R}$, which pumps magnons via Zeeman interactions \cite{Dzyapko2014,chumak2009bose},
\begin{equation}
    \hat{H}_{\text{p}}=
    \sum_{\bf k} \qty(g_{\bf k}\hat{c}^{\dagger}_{\bf k}+g_{\bf k}^*\hat{c}_{\bf k})\label{eq:classicalpump_main}
\end{equation}
with a coupling constant  $g_{\bf k}$ between magnons and the external microwave field. We have assumed in Eq. \eqref{eq:classicalpump_main} a strong coherent field, allowing us to replace the resonator operators $\hat{R}$ and $\hat{R}^{\dagger}$ with the complex amplitudes $R e^{-\ii\omega_{\rm p}t}$ and $R e^{\ii\omega_{\rm p}t}$, respectively, which have been merged into $g_{\bf k}$, as detailed in the Supplemental Material (SM). This is a natural choice in quantum optics, such as in resonance fluorescence and cavity driving \cite{scully1999quantum, breuer2002theory}. 

Prior work \cite{dzyapko2007quasiequilibrium, demidov2017chemical, demokritov2008quantum} has suggested that external pumping can increase chemical potential from zero to the energy of the lowest state.  We use a non-equilibrium open quantum systems description of magnons, deviating from the traditional equilibrium thermodynamics approaches \cite{bunkov2013spin}. This helps us derive the magnon spectral distribution directly from the Lindbladian. Single magnons are linearly one-to-one damped into single bath phonons, and magnon occupation nonlinearly hops from one at $\vb{k}$ to another at $\vb{k}'\neq \vb{k}$, mediated by a phonon. Working within this OQS framework, we substitute the collapse operators into a Lindblad master equation \cite{breuer2002theory}, describing unitary dynamics and Markovian interaction with environmental degrees-of-freedom, given by ($\hbar=1$) 
\begin{align}
\frac{\mathrm{d} \hat{\rho}_{\mathrm{S}}}{\mathrm{d} t} = & -\ii [\hat{H}_{\rm S}, \hat{\rho}_{\rm S}] \nonumber \\
& + \textstyle \sum_{\nu} \Gamma_{\nu} \qty(\hat{C}_{\nu} \hat{\rho}_{\rm S} \hat{C}_{\nu}^{\dagger} 
- \frac{1}{2} \hat{C}_{\nu}^{\dagger} \hat{C}_{\nu} \hat{\rho}_{\rm S} 
- \frac{1}{2} \hat{\rho}_{\rm S} \hat{C}_{\nu}^{\dagger} \hat{C}_{\nu}),
\end{align}
where $\hat{H}_{\rm S}$ is the internal Hamiltonian of the OQS, $\hat{\rho}_{\rm S}(t)$ is its density operator obtained by tracing out the environmental degrees-of-freedom, and $\hat{C}_{\nu}$ is the so-called collapse operator, leading to non-unitary dissipation in rate $\Gamma_{\nu}$ \cite{breuer2002theory}. The internal Hamiltonian is Eq. \eqref{eq:free-magnon}. As we do not take into account magnon-magnon interactions \cite{mag08}, the collapse operators of interest 
$\hat{C}_{\nu}\in\{\hat{c}_{\bf k}, \hat{c}_{\bf k}^{\dagger}, \hat{c}_{\bf k}^{\dagger}\hat{c}_{\bf k'}, \hat{c}_{\bf k}\hat{c}_{\bf k'}^{\dagger}\}$ are from the magnon-phonon \cite{metzger2024magnon} interactions \eqref{eq:magnonphononinterac} and the pumping Hamiltonian Eq.\eqref{eq:classicalpump_main}. The dissipation rate $\Gamma_{\nu}$ can be obtained from correlation functions of bath operators $\expval*{\hat{B}_{\nu}^{\dagger}(s)\hat{B}_{\nu}(0)}$, where $\hat{B}_{\nu}\in\{\hat{b}_{\bf k\sigma},\hat{b}_{\bf k\sigma}^{\dagger},\hat{b}_{\bf k-k'\sigma},\hat{b}_{\bf k-k'\sigma}^{\dagger}\}, $ and $\hat{B}_{\nu}(t)=e^{\ii\hat{H}_0t}\hat{B}_{\nu}e^{-\ii\hat{H}_0t}$ is their interaction picture representation \cite{breuer2002theory}. From this we compute the time evolution of the expectations of the magnon number operators, $\dv*{\expval*{\hat{n}_{\bf k}}}{t}=\tr{\hat{n}_{\bf k}\dv*{\hat{\rho}_{\rm S}}{t}}$. The resulting rate equation turns out to be very similar to the Fr\"{o}hlich rate equation,
\begin{align}
\begin{split}
    &\dv{}{t}\langle\hat{n}_{\bf k}\rangle=s_{\bf k}+\varphi_{\bf k}\left[\langle\hat{n}_{\bf k}\rangle+1-\langle\hat{n}_{\bf k}\rangle e^{\beta\omega_{\bf k}}\right]\\
    &+\sum_{\bf k'\neq k}\Lambda_{\bf kk'}\left[\langle(\hat{n}_{\bf k}+\hat{1})\hat{n}_{\bf k'}\rangle-\langle\hat{n}_{\bf k}(\hat{n}_{\bf k'}+\hat{1})\rangle e^{\beta(\omega_{\bf k}-\omega_{\bf k'})}\right],
\end{split}\label{eq:quantumFrohlich}
\end{align}
where $s_{\bf k} = 2\pi \abs{g_{\bf k}}^2$, $\varphi_{\bf k}=2\pi |V_{\sigma{\bf k}}|^2\bar{n}_{\rm B}(\omega_{\bf k})$, $\Lambda_{\bf kk'}=2\pi |V_{\sigma{\bf kk'}}|^2\bar{n}_{\rm B}(\omega_{\bf k}-\omega_{\bf k'})$, and $\bar{n}_{\rm B}(\omega)=1/(e^{\beta \omega}-1)$ is the mean number of phonons in a thermally occupied mode of frequency $\omega$. Note that the above couplings, $s_{\bf k}$, $\varphi_{\bf k}$, and $\Lambda_{\bf kk'}$, should be weak, compared to the magnon spectral frequency $\omega_{\bf k}$ (see SM for details).

Fr\"ohlich hypothesized uncorrelated occupations to have an exact stationary solution,
\begin{align}
    \expval{n_k}&=\qty(1+\frac{s_k}{\varphi_k+\sum_{j\neq k}\Lambda_{kj}\expval{n_j}})\frac{1}{A_ke^{\beta\omega_k}-1},
    \label{eq:condensate}
\end{align}
where $A_k=\frac{\varphi_k+\sum_{j\neq k}\Lambda_{kj}(1+\expval{n_j})e^{-\beta\omega_j}}{\varphi_k+\sum_{j\neq k}\Lambda_{kj}\expval{n_j}}$,
showing that the mean number distribution acquires the Bose-Einstein-like form \cite{frohlich1968bose,frohlich1968long}. Eq. (\ref{eq:condensate}) shows that $\expval{n_k}$ is linearly proportional to $s_k$ when $\varphi_k$ and $\Lambda_{kj}$ are fixed. When $s_k$ is zero, the nonlinear terms vanish ($A_k=1$), and the system stays in the Planck distribution. While originally Fr\"ohlich condensation was derived phenomenologically, it has also been re-derived by leveraging a classical master equation of birth-and-death processes in a system of harmonic oscillators with a discrete spectrum of frequencies $\{\omega_1,\ldots, \omega_L\}$ and occupation number $n_k$ \cite{preto2017semi}, 
\begin{equation}
\begin{split}
    &\dv{t}\expval{n_k}=s_k+\varphi_k\left(\expval{n_k}+1-\expval{n_k}e^{\beta\omega_k}\right)\\
    &+\sum_{j\neq k}^L\Lambda_{kj}\qty[\qty(\expval{n_kn_j}+\expval{n_j})-\left(\expval{n_kn_j}+\expval{n_k}\right)e^{\beta(\omega_k-\omega_j)}].
\end{split}
    \label{eq:pretoFrohlich}
\end{equation}
where, as typical for classical stochastic processes, $\langle\cdots\rangle$ means averaging over stochastic trajectories. The derived rate equation \eqref{eq:pretoFrohlich} accounts for possible classical correlations between different occupation numbers $\expval{n_kn_j}$; while the original Fr\"ohlich rate equation inherently imposes vanishing correlations, i.e.\ postulates $\expval{n_kn_j} = \expval{n_k}\expval{n_j}$.  The uncorrelated approximation works well in this semi-classical model for oscillators of THz frequencies at ambient temperatures \cite{preto2017semi}. 
=
\begin{figure}[b]
    \centering
    \includegraphics[width=0.92\linewidth]{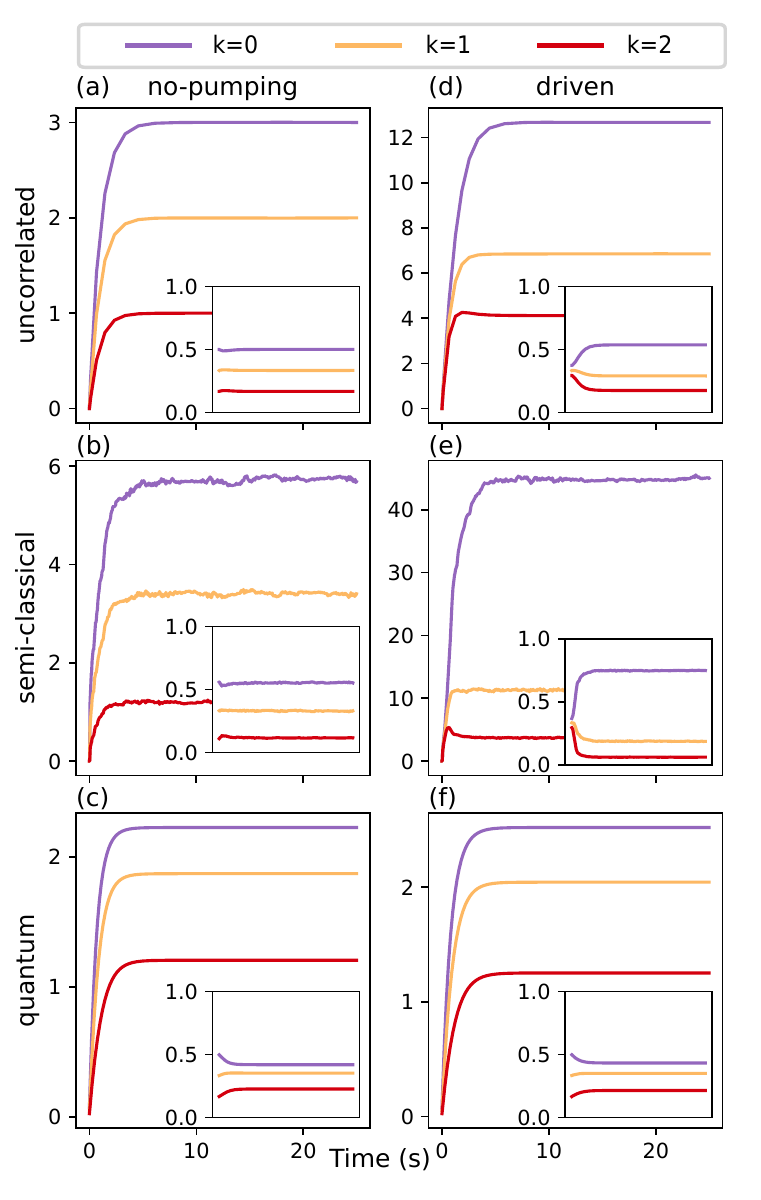}
    \caption{Occupation number distribution for a three-mode toy system for original uncorrelated, semi-classical (\ref{eq:pretoFrohlich}), and quantum (\ref{eq:quantumFrohlich}) formulations. The first row of subplots corresponds to the case without pumping $s_k=0$, and the second row is for the case with a time-dependent pumping. The inset plot in each subplot is the corresponding normalized occupation: $\expval{n_k}/\sum_k \expval{n_k}$ or $\expval{\hat{n}_{\bf k}}/\sum_{\bf k} \expval{\hat{n}_{\bf k}}$ (Here, ${\bf k}$ and $k$ are equivalent indexing of energy levels in the toy system).}
    \label{fig:occupation}
\end{figure}
A similar Bose-Einstein-like distribution, $\expval{\hat{n}_{\bf k}}\sim 1/(A_{\bf k}\exp(\beta\omega_{\bf k})-1)$, is appealing to derive using Eq. \eqref{eq:quantumFrohlich}. There are previous attempts \cite{wu1977bose, wu1978bose,wang2022full} that reached Eq. \eqref{eq:quantumFrohlich}; however, they derived the pumping term $s_{\bf k}$ through an additional linear transition and required the source field of pumping at infinite temperature. In our derivation, we avoid such an unrealistic context of infinite temperature for the pumping field, $\beta_{\rm p}\to0$, by considering a coherent electromagnetic (EM) field pumping the magnons classically.
\begin{figure*}
    \centering
    \includegraphics[width=0.95\textwidth]{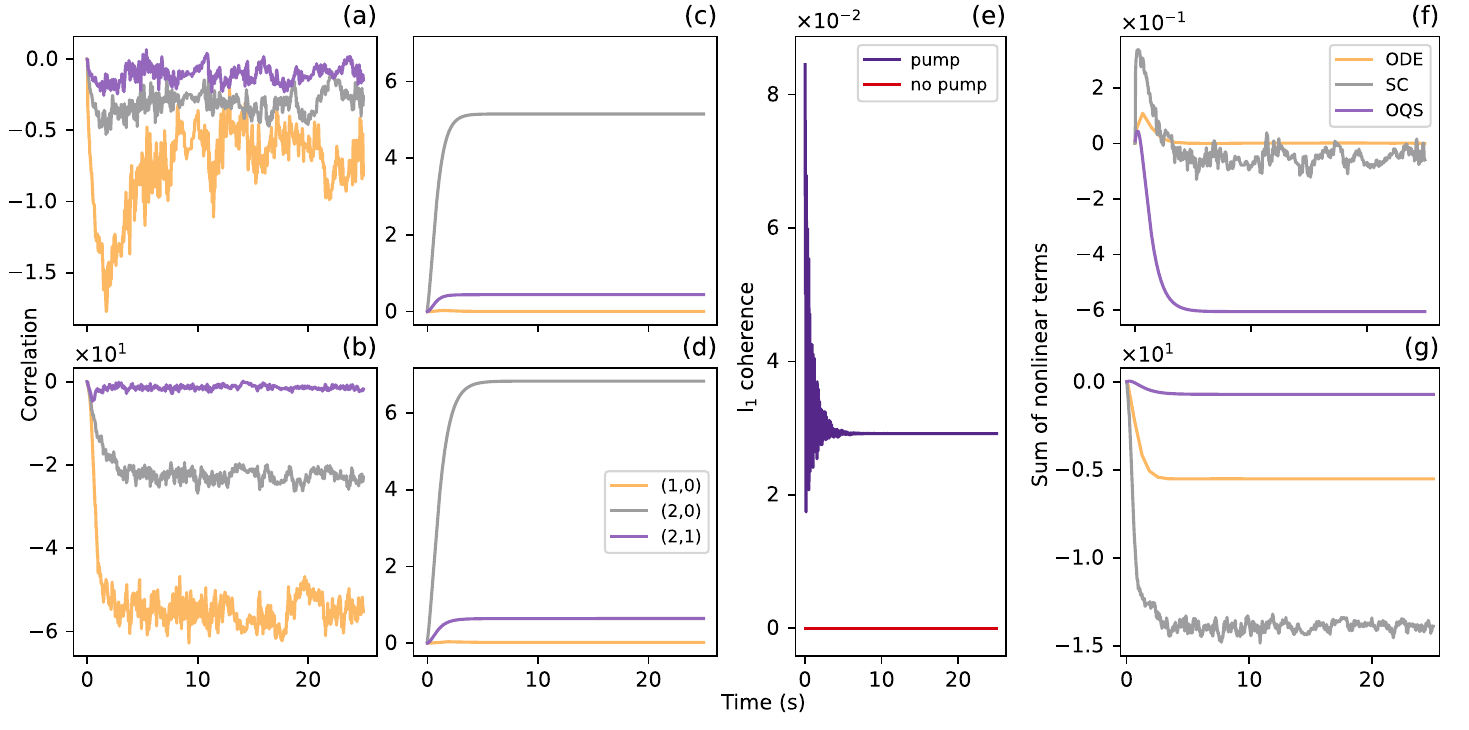}
    \caption{Comparison of correlations, quantum coherence, and nonlinear terms across frameworks. Correlations are presented in (a and b) for the semi-classical framework and in (c and d) for the OQS framework. $l_1$ quantum coherence calculations are also included in (e). (f and g) illustrate the sum of nonlinear terms in the Fr\"{o}hlich rate equations across the three frameworks. The top row of subplots corresponds to scenarios without pumping, while the bottom row depicts cases with pumping. Parameters are consistent with those used in Fig. (\ref{fig:occupation}).}
    \label{fig:compare}
\end{figure*}
We consider three similar, but by no means equivalent, versions of the Fr\"{o}hlich rate equation: Eq. \eqref{eq:quantumFrohlich}, which we derived from the magnon-phonon Hamiltonian in the OQS paradigm, Eq. \eqref{eq:pretoFrohlich}, derived previously \cite{preto2017semi} in the semi-classical picture; and the one employing the uncorrelated approximation in Eq. \eqref{eq:condensate}, as originally hypothesized by Fr\"ohlich \cite{preto2017semi}. The key difference between them is the correlations between occupations of different modes (namely different wave numbers). The uncorrelated approximation renders two-point correlation functions trivial
\begin{subequations}
    \begin{eqnarray}
        r_{\text{SC}}&=&\expval{n_k n_j}-\expval{n_k}\expval{n_j} \approx 0\label{eq:scfactorized}\\
        r_{\text{OQS}}&=&\expval{\hat{n}_{\bf k} \hat{n}_{\bf k'}}-\expval{\hat{n}_{\bf k}}\expval{\hat{n}_{\bf k'}}\approx 0.\label{eq:oqsfactorized}
    \end{eqnarray}\label{eq:factorized}
\end{subequations}
By restricting ourselves to the first three low-lying spectral states for illustrative purposes, we avoid the need for costly numerical simulation of a large OQS \cite{purkayastha2024difference, chowdhury2024, villanueva2024stochastic, huijgen2024training, driv22, smith2019quantum} with many states and large occupation numbers. This is sufficient to allow us to challenge the validity of the uncorrelated \cite{wang2022full} approximation \eqref{eq:factorized}, and show that correlations can significantly alter the spectral distribution over magnon modes. We impose equilibrium occupations for each mode: $n_0 = 3$, $n_1 = 2$, and $n_2 = 1$, at $\beta=0.0033$ (corresponding to 300K). The labeling $k=0$ represents the ground state, $k=1$ represents the first-excited state, and $k=2$ represents the highest energy state. Other parameters can be found in Table.\ \ref{tab:1}. 
\begin{table}[t]
\caption{\label{tab:1} Parameters used for numerical simulations in this work. For the purpose of simplifying calculations, $\hbar=k_B=1$ is set. The original uncorrelated Fr\"{o}hlich rate equation is simulated by \texttt{scipy.solve\_ivp} in \texttt{Python}, the semi-classical one can be simulated by Gillespie algorithm \cite{preto2017semi}, and the OQS simulation uses \texttt{Qutip} in \texttt{Python}.}
\begin{ruledtabular}
\begin{tabular}{ccccccc}
 $n_0$ &$n_1$ & $n_2$ & $\beta$ &$g_{\bf k}$ &$\abs{V_{\bf \sigma k}}^2$&$\abs{V_{\bf \sigma kk'}}^2$\\
\hline
3&2&1&0.0033& $e^{\ii\pi t}$ & $1/2\pi$ & $0.05/2\pi$\\
\end{tabular}
\end{ruledtabular}
\end{table}

Fig. (\ref{fig:occupation}) displays different steady-state occupation distributions among the three rate equations. The first column (a, b, and c) corresponds to the case without pumping. The uncorrelated case consistently presents preset equilibrium Planck distribution. The occupations are slightly lower in the OQS case and higher in the semi-classical one (the unexpectedly large energy spacings in latter could be attributed to limitations in the semi-classical treatment of a three mode model). Also, their spacing between occupations at each level are narrower in (c) and wider in (b), compared to the results in (a). In the presence of pumping, as illustrated in the second column (d, e, and f), the degree of condensation varies. The semi-classical result in (e) exhibits the most significant condensation: the ground state is dominantly occupied, followed by the uncorrelated case in (d). The condensation effects can be further observed by comparing the normalized distributions, as shown in the insets of (d) and (e), in the presence of pumping to those in the absence of pumping. The proportionality of ground-state occupation to total occupation is highest in the semi-classical model, followed by the uncorrelated one. The OQS result shows minimal condensation: the ground state is not predominantly occupied, although there is a proportional increase in its expected occupation, as shown in the normalized distribution in the inset of (f) compared to that of (c).
\par 
Fig. (\ref{fig:compare}) further compares nonlinear terms and correlations among the three models, and shows the coherence induced by pumping in Eq. \eqref{eq:quantumFrohlich}. Without pumping, Figs. (\ref{fig:compare}.f and \ref{fig:compare}.g) show that the nonlinear terms in the original uncorrelated model sum to zero, while in the steady states of both the semi-classical and OQS results, the sums become nonzero. In the presence of pumping, these sums continue to diverge from the results of the uncorrelated case. To gain insights, we demonstrate a breakdown of the uncorrelated approximation \eqref{eq:factorized} by examining $r_{\rm SC}$ in Figs. (\ref{fig:compare}.a and \ref{fig:compare}.b) and $r_{\rm OQS}$ in Figs. (\ref{fig:compare}.c and \ref{fig:compare}.d). The classical correlations are numerically calculated by averaging the product $n_kn_j$ over 1000 trajectories from the Monte Carlo sampling of the Gillespie algorithm \cite{preto2017semi}, and the quantum correlations are calculated by taking the expectation of the $\hat{n}_{\bf k} \hat{n}_{\bf k'}$ operators with respect to the magnon density matrix in \texttt{Qutip}. We observe that $r_{\rm SC}$ is non-positive, which implies anti-correlations leading to spacing between occupations at different energy levels to be wider. Conversely, $r_{\rm OQS}$ is non-negative, suggesting that quantum correlations narrow the spacing. In particular, the quantum correlation between the highest-excited and ground states (2,0) is largest, which results in such a minimal condensation. 
\par
Compared to the uncorrelated case, the nontrivial classical and quantum correlations explain why the occupation distributions differ significantly among the three models in Fig. (\ref{fig:occupation}). To assess whether the pumping \eqref{eq:classicalpump_main} induces quantum coherence and thereby affects condensation, we examine the $l_1$ coherence $C_{l_1} = \sum_{i\neq j}\abs{\rho_{ij}}$ in Fig. (\ref{fig:compare}.e). $C_{l_1}$ sums the magnitudes of all off-diagonal density matrix elements. In the absence of pumping, quantum coherence remains zero $C_{l_1}=0$, but quantum correlations are non-trivial, with a nonzero sum of nonlinear terms still present. In the presence of pumping, the quantum coherence appears $C_{l_1}>0$, and the quantum correlations also increase. It is difficult to identify whether this increase in quantum correlations is due to quantum coherence or the pumping affecting the diagonal density matrix elements, as classical anti-correlations are also amplified under pumping, as shown in (b). But the key factor contributing to the difference in occupation distribution between the OQS \eqref{eq:quantumFrohlich} and the uncorrelated case appears to be the quantum correlations, regardless of pumping, rather than coherence. In principle, Eq. \eqref{eq:quantumFrohlich} can be transformed to reflect the semi-classical rate equation \eqref{eq:pretoFrohlich}, if only the diagonal elements of the density matrix are taken into account, with the quantum correlations then becoming equivalent to classical ones. However, our results imply a possible over-prediction of condensation, as for both the semi-classical and original rate equations we have numerically shown condensation to be more pronounced and to undergo a sharper second-order phase transition with respect to pumping power at the thermodynamic limit \cite{preto2017semi, nardecchia2018out}. 

The emergence of BECs \cite{ketterle1997coherence, Gomonay2024, anderson1995observation, davis1995bose, ketterle1997coherence} generally requires accessing quantum matter at extremes. Nevertheless, recent observations  \cite{hakala2018bose, klaers2010bose, balili2007bose, kasprzak2006bose, deng2002condensation} show a more feasible alternative by demonstrating condensates in driven quasi-equilibrium systems, like in the case of magnons \cite{demokritov2006bose, serga2014bose}. However, the non-equilibrium mechanism underlying such room-temperature condensations remains uncertain \cite{PhysRevResearch.5.L022065}. In our derivation, the resulting rate equation of expected magnon occupation \eqref{eq:quantumFrohlich} is similar to the one governing the Fr\"{o}hlich condensation, differing mainly with respect to the uncorrelated approximation \eqref{eq:oqsfactorized}. Our three-mode system simulations show that at room temperature, the BEC-like condensation is not obvious, since the quantum correlations $r_{\rm OQS}>0$ narrow the spacing between occupations on each energy level. Compared to the semi-classical model of Fr\"{o}hlich condensation \eqref{eq:pretoFrohlich}, we note that the classical anti-correlations $r_{\rm SC}<0$ lead to a more condensed distribution in the ground state. This suggests that non-trivial correlations neglected in previous studies \cite{wang2022full,wu1977bose,wu1978bose} should be treated more carefully to avoid over-predictions. The concurrence of both the predicted phenomena and theoretical formulations of Fr\"{o}hlich condensation with those of magnon condensation suggests intricate connections. This is reflected by their common features: condensed, steady ground states necessitate nonlinear interactions and the emergence is thresholded by external pumping \cite{demokritov2006bose, frohlich1968long}. 

By interpolating quantum and classical regimes, our treatment offers a natural framework for characterizing condensation in pumped bosonic systems. While our treatment is intentionally simplified to extract the effects of magnon-phonon interactions and non-trivial correlations, such that the complete physical characteristics of actual magnon condensation are not necessarily captured, it is adaptable and generalizable beyond the three-mode example. It is possible that different system parameters could vary the signs of classical and quantum correlations, and thus lead to different condensation effects. Although constrained by computational bottlenecks in OQS simulations for realistic models \cite{tutorialOQS} at the thermodynamic limit for finite mode numbers--beyond just three modes, we anticipate that our investigation will guide the inclusion of non-trivial correlations in future studies involving magnon-magnon interactions and larger systems, permitting an assessment of the limits these might place on condensation in experimental situations. For instance, one could apply more sophisticated approximations, such as Wigner function approaches \cite{wigner19, wigner21magnon}, on top of our OQS formulation, to eventually simulate larger systems of realistic complexity. Nevertheless, our results suggest a superior paradigm of high-temperature condensation in pumped bosonic systems compared to the traditional treatments based on equilibrium thermodynamics, with our three-mode example sufficiently reflecting the general characteristics of high-temperature condensation. The next step will be the application of this new paradigm to the more complicated case of longitudinal (that is, two-boson) pumping which we keep for future research. 

\begin{acknowledgments}
\textit{Acknowledgments}.--- We acknowledge use of University of Exeter's HPC facility. W.X. thanks Dr. Eduardo Domínguez and Miao Hu for their helpful discussion and comments on the theory and acknowledges the financial support from an Erasmus+ traineeship Grant. F.T.C., L.D.S., and D.R.K. acknowledge support from the Office of Naval Research (ONR Award No. N62909-21-1-2018) and the Engineering and Physical Sciences Research Council (EP/X027376/1). 
\end{acknowledgments}

\nocite{*}

\providecommand{\noopsort}[1]{}\providecommand{\singleletter}[1]{#1}%

\setcounter{equation}{0}
\setcounter{figure}{0}
\renewcommand{\theequation}{E\arabic{equation}}
\renewcommand{\thefigure}{A\arabic{figure}}

\clearpage
\onecolumngrid 
\textheight=679pt
\fontsize{10}{14}\selectfont 
\setcounter{figure}{0} \newpage
\setcounter{equation}{0}
\renewcommand{\theequation}{S\arabic{equation}}
\renewcommand{\thefigure}{S\arabic{figure}}
\begin{center}
\textbf{\large Supplemental Material}:\\ \Large{Fr\"{o}hlich versus Bose-Einstein Condensation in Pumped Bosonic Systems}
\end{center}

\section{Derivation of Magnon Hamiltonians}\label{append:magnon}
\subsection{Dipolar Sums}\label{append:dipolarsum}
The dipolar sums will be used in the following expansions by Holstein-Primakoff (HP) and Dyson-Maleev (DM) transformations. The three-dimensional dipolar sums, approximated for wave vectors $kd\ll1$, are given by \cite{s_kreisel2009microscopic, s_li2013phase}:
\begin{align}
    D_1({\bf k})=&\frac{a_0^3}{N}\sum_{j,j'}\frac{1-3\hat{z}_{jj'}^2}{r_{jj'}^3}\exp(\ii{\bf k}\cdot{\bf r}_{jj'})= -4\pi\left[\frac{1}{3}+\cos^2(\theta_{\bf k})(F_{\bf k}-1)\right],\\
    D_2({\bf k})=&\frac{a_0^3}{N}\sum_{j,j'}\frac{(\hat{x}_{jj'}-\ii\hat{y}_{jj'})^2}{r_{jj'}^3}\exp(\ii{\bf k}\cdot{\bf r}_{jj'})= -\frac{4\pi}{3}\left[F_{\bf k}+\sin^2(\theta_{\bf k})(F_{\bf k}-1)\right],\\
    D_3({\bf k})=&\frac{a_0^3}{N}\sum_{j,j'}\frac{(\hat{x}_{jj'}-\ii\hat{y}_{jj'})\hat{z}_{jj'}}{r_{jj'}^3}\exp(\ii{\bf k}\cdot{\bf r}_{jj'})=\frac{4\pi}{3}(F_{\bf k}-1)\sin{\theta_{\bf k}}\cos{\theta_{\bf k}}.
\end{align}
where $F_{\bf k}=\qty(1-e^{-\abs{\vb{k}}d})/\abs{\vb{k}}d$ and $d$ denotes the thickness of the film \cite{s_tupitsyn2008stability}.
\subsection{Holstein-Primakoff and Dyson-Maleev Transformation}
The physical properties of YIG films can be described by the Heisenberg-like quantum ferromagnet model on a cubic lattice \cite{s_rezende2009theory, s_akhiezer1968spin}, which contain both exchange and dipole-dipole interactions between localized spins, and the Zeeman interaction with an external magnetic field:
\begin{align}
    \begin{split}
        \hat{H}&=-\sum_{j,j'}J_{jj'}\hat{\mathbf{S}}_j\cdot\hat{\mathbf{S}}_{j'}-\gamma{\bf H}_0\cdot\sum_j\hat{\bf S}_j-\frac{\gamma^2}{2}\sum_{j,j'}\frac{1}{r_{jj'}^3}\left[3(\hat{\mathbf{S}}_j\cdot\hat{\mathbf{r}}_{jj'})(\hat{\mathbf{S}}_{j'}\cdot\hat{\mathbf{r}}_{jj'})-\hat{\mathbf{S}}_j\cdot\hat{\mathbf{S}}_{j'}\right],
        \label{eq:quantumferroHami}
    \end{split}
\end{align}
where the sums are taken over the lattice sites denoted by $\mathbf{r}_{j}$, and $\hat{\mathbf{r}}_{jj'}=\mathbf{r}_{jj'}/r_{jj'}$ is a unit vector along the direction of $\mathbf{r}_{jj'}=\mathbf{r}_j-\mathbf{r}_{j'}$. Here, $\gamma$ is the gyromagnetic ratio, and ${\bf H}_0$ is the external magnetic field.
\par 
The HP and DM transformations are equivalent under third-order expansions of the Hamiltonian \eqref{eq:quantumferroHami}. The only differences are: (1) HP leads to an infinite series of expansions, while DM only results in up to sixth-order terms; (2) HP leads to a Hermitian Hamiltonian, while DM leads to a non-Hermitian Hamiltonian beyond third-order expansions; (3) magnon oscillators are ideal oscillators in the DM, while they have an upper bound of occupations in the HP. If truncated at the third-order, they are equivalent. Here, we will follow the DM transformation but approximated to the second-order expansion at large-$S$ limit ($S\gg \hat{a}_j^{\dagger}\hat{a}_j$), which is equivalent to the HP one.
\begin{subequations}
    \begin{eqnarray}
    \hat{S}_j^+&=& \sqrt{2S}\left(1-\frac{\hat{a}_j^{\dagger}\hat{a}_j}{2S}\right)\hat{a}_j\approx\sqrt{2S}\hat{a}_j\\
    \hat{S}_j^-&=&\sqrt{2S}\hat{a}_j^{\dagger}\\
    \hat{S}_j^z&=&S-\hat{a}_j^{\dagger}\hat{a}_j
\label{eq:HPDMexpansion}
\end{eqnarray}
\end{subequations}
The reason is that DM leads to ideal magnon oscillators; while the HP upper bounds the highest occupation of the oscillators, $\hat{a}_j^{\dagger}\hat{a}_j\in[0,2S]$. For a general theory of high-temperature condensation, we wish to start with the ideal oscillators and thus adopt the DM description. Thus for the Zeeman term, we have
\begin{align}
    \hat{H}_{\text{Z}}=&-g\mu_B\sum_j\hat{S}_j^zH_0^z=-g\mu_BH_0^zNS+g\mu_BH_0^z\sum_j\hat{a}_j^{\dagger}\hat{a}_j.
\end{align}
The same is taken for the exchange interaction term;
\begin{align}
    \hat{H}_{\text{ex}}=&-\frac{J}{2}\sum_{j,\boldsymbol{\delta}}\left(\hat{S}_j^+\hat{S}_{j+\boldsymbol{\delta}}^-+\hat{S}_j^-\hat{S}_{j+\boldsymbol{\delta}}^++2\hat{S}_j^z\hat{S}_{j+\boldsymbol{\delta}}^z\right)\\
    \approx&-\frac{J}{2}\sum_{j,\boldsymbol{\delta}}\left[2S\hat{a}_j\hat{a}_{j+\boldsymbol{\delta}}^{\dagger}+2S\hat{a}_j^{\dagger}\hat{a}_{j+\boldsymbol{\delta}}+2\left(S^2-S\hat{n}_{j+\boldsymbol{\delta}}-S\hat{n}_j\right)\right]
\end{align}
where $\vb*{\delta}$ denotes the vector to the nearest neighboring sites. Note that the above substitutions have been truncated at the second order; namely, 
\begin{align}
    \hat{S}_j^+\hat{S}_{j'}^-&=2S\hat{a}_j\hat{a}_{j'}^{\dagger}&
    \hat{S}_j^-\hat{S}_{j'}^+&=2S\hat{a}_j^{\dagger}\hat{a}_{j'}&
    \hat{S}_j^z\hat{S}_{j'}^z&=S^2-S\hat{n}_{j'}-S\hat{n}_j.
\end{align}
Written in an ascending order, we have
\begin{align}
    \hat{H}_{\text{ex}}=&-NZ_0JS^2-JS\sum_{j,j'}\left(\hat{a}_j\hat{a}_{j'}^{\dagger}+\hat{a}_j^{\dagger}\hat{a}_{j'}-\hat{a}_{j'}^{\dagger}\hat{a}_{j'}-\hat{a}_j^{\dagger}\hat{a}_j\right).
\end{align}
For the dipolar term, it can be written out as
\begin{align}
    \begin{split}
        \hat{H}_{\text{dip}} &= \frac{1}{2}(g\mu_B)^2\sum_{j,j'} \left\{ \frac{1}{r_{jj'}^3} \left[ \frac{1}{2}\left( \hat{S}_j^+ \hat{S}_{j'}^- + \hat{S}_j^- \hat{S}_{j'}^+ \right) + \hat{S}_j^z \hat{S}_{j'}^z \right] \right. \\
        &\quad - \frac{3}{r_{jj'}^5} \left[ z_{jj'}^2 \hat{S}_j^z \hat{S}_{j'}^z + \frac{1}{4} r_{jj'}^+ r_{jj'}^- \left( \hat{S}_j^+ \hat{S}_{j'}^- + \hat{S}_j^- \hat{S}_{j'}^+ \right) \right. \\
        &\qquad \left. \left. + \frac{1}{4} \left( r_{jj'}^- \right)^2 \hat{S}_j^+ \hat{S}_{j'}^+ + \frac{1}{4} \left( r_{jj'}^+ \right)^2 \hat{S}_j^- \hat{S}_{j'}^- + \frac{1}{2} z_{jj'} r_{jj'}^- \left( \hat{S}_j^z \hat{S}_{j'}^+ + \hat{S}_j^+ \hat{S}_{j'}^z \right) + \frac{1}{2} z_{jj'} r_{jj'}^+ \left( \hat{S}_j^z \hat{S}_{j'}^- + \hat{S}_j^- \hat{S}_{j'}^z \right) \right] \right\}
    \end{split}
\end{align}
where $r_{jj'}^{\pm}=x_{jj'}\pm \ii y_{jj'}$. The operator multiplications can be summarized as follows:
\begin{align}
    \hat{S}_{j}^z\hat{S}_{j'}^+&= S\sqrt{2S}\hat{a}_{j'}-\sqrt{\frac{S}{8}}\hat{n}_{j'}\hat{a}_{j'}-\sqrt{2S}\hat{n}_j\hat{a}_{j'};\\
    \hat{S}_{j}^z\hat{S}_{j'}^-&= S\sqrt{2S}\hat{a}_{j'}^{\dagger}-\sqrt{\frac{S}{8}}\hat{a}_{j'}^{\dagger}\hat{n}_{j'}-\sqrt{2S}\hat{n}_j\hat{a}_{j'}^{\dagger};\\
    \hat{S}_{j}^+\hat{S}_{j'}^z&= S\sqrt{2S}\hat{a}_{j}-\sqrt{\frac{S}{8}}\hat{n}_{j}\hat{a}_{j}-\sqrt{2S}\hat{a}_{j}\hat{n}_{j'};\\
    \hat{S}_{j}^-\hat{S}_{j'}^z&= S\sqrt{2S}\hat{a}_{j}^{\dagger}-\sqrt{\frac{S}{8}}\hat{a}_{j}^{\dagger}\hat{n}_{j}-\sqrt{2S}\hat{a}_{j}^{\dagger}\hat{n}_{j'};\\
    \hat{S}_j^+\hat{S}_{j'}^+&=2S\hat{a}_j\hat{a}_{j'};\\
    \hat{S}_j^-\hat{S}_{j'}^-&=2S\hat{a}_j^{\dagger}\hat{a}_{j'}^{\dagger}.
\end{align}
Again, by truncating the expansion at the second order, the dipolar term reads
\begin{align}
    \begin{split}
        \hat{H}_{\text{dip}} =& \frac{1}{2}(g\mu_B)^2\sum_{j,j'} \Biggl\{ \frac{1}{r_{jj'}^3} \Biggl[ \frac{1}{2}\left( 2S\hat{a}_j\hat{a}_{j'}^{\dagger} + 2S\hat{a}_j^{\dagger}\hat{a}_{j'} \right) + S^2 - S\hat{n}_{j'} - S\hat{n}_j \Biggr] \\
        & - \frac{3}{r_{jj'}^5} \Biggl[ z_{jj'}^2\left( S^2 - S\hat{n}_{j'} - S\hat{n}_j \right) \\
        & + \frac{1}{4} r_{jj'}^+ r_{jj'}^- \left( 2S\hat{a}_j\hat{a}_{j'}^{\dagger} + 2S\hat{a}_j^{\dagger}\hat{a}_{j'} \right) + \frac{1}{4}\left( r_{jj'}^- \right)^2 2S\hat{a}_j\hat{a}_{j'} + \frac{1}{4}\left( r_{jj'}^+ \right)^2 2S\hat{a}_j^{\dagger}\hat{a}_{j'}^{\dagger} \\
        & + \frac{1}{2} z_{jj'} r_{jj'}^- \left( S\sqrt{2S}\hat{a}_{j'} + S\sqrt{2S}\hat{a}_j \right) + \frac{1}{2} z_{jj'} r_{jj'}^+ \left( S\sqrt{2S}\hat{a}_{j'}^{\dagger} + S\sqrt{2S}\hat{a}_j^{\dagger} \right) \Biggr] \Biggr\}
    \end{split}
\end{align}

Written in an ascending order, the lower-order terms read
\begin{align}
    \hat{H}_{\text{dip}}^{(0)} =& \frac{1}{2}(g\mu_B)^2\sum_{j,j'}\left( \frac{1}{r_{jj'}^3}S^2 - \frac{3}{r_{jj'}^5}z_{jj'}^2S^2 \right) = \frac{1}{2}(g\mu_B)^2 N S^2 \sum_{r} \left( \frac{1}{r^3} - \frac{3 z^2}{r^5} \right) \\
    \hat{H}_{\text{dip}}^{(1)} =& -\frac{1}{2}(g\mu_B)^2 S \sqrt{2S} \sum_{j} \sum_{j' \neq j} \frac{3 z_{jj'}}{2 r_{jj'}^5} \left[ r_{jj'}^- \left( \hat{a}_{j'} + \hat{a}_{j} \right) + r_{jj'}^+ \left( \hat{a}_{j'}^{\dagger} + \hat{a}_{j}^{\dagger} \right) \right] \\
    \begin{split}
    \hat{H}_{\text{dip}}^{(2)} =& \frac{1}{2}(g\mu_B)^2 S \sum_{j,j'} \Biggl\{ \frac{1}{r_{jj'}^3} \left( \hat{a}_j \hat{a}_{j'}^{\dagger} + \hat{a}_j^{\dagger} \hat{a}_{j'} - \hat{a}_{j'}^{\dagger} \hat{a}_{j'} - \hat{a}_{j}^{\dagger} \hat{a}_{j} \right) \\
    & - \frac{3}{r_{jj'}^5} \Biggl[ -z_{jj'}^2 \left( \hat{a}_{j'}^{\dagger} \hat{a}_{j'} + \hat{a}_{j}^{\dagger} \hat{a}_{j} \right) + \frac{1}{2} r_{jj'}^+ r_{jj'}^- \left( \hat{a}_j \hat{a}_{j'}^{\dagger} + \hat{a}_j^{\dagger} \hat{a}_{j'} \right) \\
    & \quad + \frac{1}{2} \left( r_{jj'}^- \right)^2 \hat{a}_j \hat{a}_{j'} + \frac{1}{2} \left( r_{jj'}^+ \right)^2 \hat{a}_j^{\dagger} \hat{a}_{j'}^{\dagger} \Biggr] \Biggr\} \\
    =& \frac{1}{2}(g\mu_B)^2 S \sum_{j,j'} \frac{1}{r_{jj'}^3} \Biggl[ -\left( 1 - \frac{3 z_{jj'}^2}{r_{jj'}^2} \right) \hat{a}_j^{\dagger} \hat{a}_j - \left( 1 - \frac{3 z_{jj'}^2}{r_{jj'}^2} \right) \hat{a}_{j'}^{\dagger} \hat{a}_{j'} \\
    & + \left( 1 - \frac{3}{2} \frac{ r_{jj'}^+ r_{jj'}^- }{ r_{jj'}^2 } \right) \hat{a}_j^{\dagger} \hat{a}_{j'} + \left( 1 - \frac{3}{2} \frac{ r_{jj'}^+ r_{jj'}^- }{ r_{jj'}^2 } \right) \hat{a}_j \hat{a}_{j'}^{\dagger} \\
    & - \frac{3}{2} \frac{ \left( r_{jj'}^- \right)^2 }{ r_{jj'}^2 } \hat{a}_j \hat{a}_{j'} - \frac{3}{2} \frac{ \left( r_{jj'}^+ \right)^2 }{ r_{jj'}^2 } \hat{a}_j^{\dagger} \hat{a}_{j'}^{\dagger} \Biggr],
    \end{split}
\end{align}
\par
In terms of normal modes, Fourier transform can be introduced due to translational symmetry
\begin{align*}
    \hat{a}_j&=\frac{1}{\sqrt{N}}\sum_{\mathbf{k}}\hat{a}_{\mathbf{k}}e^{\ii\mathbf{k}\cdot\mathbf{r}_j}, &\hat{a}^{\dagger}_j&=\frac{1}{\sqrt{N}}\sum_{\mathbf{k}}\hat{a}^{\dagger}_{\mathbf{k}}e^{-\ii\mathbf{k}\cdot\mathbf{r}_j};
    &\hat{a}_{\mathbf{k}}&=\frac{1}{\sqrt{N}}\sum_{j}\hat{a}_{j}e^{-\ii\mathbf{k}\cdot\mathbf{r}_j},&\hat{a}_{\mathbf{k}}^{\dagger}&=\frac{1}{\sqrt{N}}\sum_{j}\hat{a}_{j}^{\dagger}e^{\ii\mathbf{k}\cdot\mathbf{r}_j}.
\end{align*}
where $N$ is the number of unit cells, and $\mathbf{k}$ lies in the first Brillouin zone \cite{s_akhiezer1968spin}. The completeness and orthogonality relation of the eigenfunctions gives $\sum_je^{\ii(\mathbf{k}-\mathbf{k}')\cdot\mathbf{r}_j}=N\delta_{\mathbf{k}\mathbf{k}'}$. The first-order Hamiltonian is vanished;
\begin{align*}
\begin{split}
    \hat{H}^{(1)}&=-\frac{3S}{4}(g\mu_B)^2\sqrt{\frac{2S}{N}}\sum_{\mathbf{k}}\sum_{j,j'}\left[\frac{z_{jj'}r_{jj'}^-}{r_{jj'}^5}\left(e^{\ii\mathbf{k}\cdot\mathbf{r}_j}+e^{\ii\mathbf{k}\cdot\mathbf{r}_{j'}}\right)\hat{a}_{\mathbf{k}}+\frac{z_{jj'}r_{jj'}^+}{r_{jj'}^5}\left(e^{-\ii\mathbf{k}\cdot\mathbf{r}_j}+e^{-\ii\mathbf{k}\cdot\mathbf{r}_{j'}}\right)\hat{a}_{\mathbf{k}}^{\dagger}\right]\\
    &=-\frac{3S}{4}(g\mu_B)^2\sqrt{2SN}\sum_{r}\left[\frac{zr^-}{r^5}a_{\mathbf{0}}+\frac{zr^+}{r^5}a_{\mathbf{0}}^{\dagger}\right]=0.
\end{split}
\end{align*}
Taking into account that $\sum_je^{\ii\mathbf{k}\cdot\mathbf{r}_j}=N\delta_{\mathbf{k}\mathbf{0}}$, we obtain
\begin{align}
    \sum_{j\neq j'}\frac{z_{jj'}r_{jj'}^{\pm}}{r_{jj'}^5}e^{\pm \ii\mathbf{k}\cdot\mathbf{r}_j}=N\sum_{r}\frac{zr^{\pm}}{r^5}\delta_{\mathbf{k}\mathbf{0}},
\end{align}
Since the $z$-axis is the symmetry axis, the sums are equal to zero,
\begin{align}
    \sum_{\mathbf{r}}\frac{zr^{\pm}}{r^5}=\sum_{\mathbf{r}}\frac{z(x\pm \ii y)}{r^5}=0.
\end{align}
If there is pumping, the magnons are pumping via Zeeman interactions,
\begin{subequations}
    \begin{eqnarray}
        \hat{H}_{\text{p}}&=&-\gamma\sqrt{\frac{SN}{2}}\sum_{\mathbf{k}}\mqty[\left(u_{\bf k}h^*+
        v_{\bf k}^*h\right)\hat{c}_{\bf k}+\text{h.c.}]\label{eq:pumpHami}\\
        &\Longrightarrow&\sum_{\bf k} \qty(g_{\bf k}\hat{c}^{\dagger}_k+g_{\bf k}^*\hat{c}_k),\label{eq:classicalpump}
    \end{eqnarray}  
\end{subequations}
where $h \sim (\hat{R}+\hat{R}^{\dagger})$ denotes the pumping field \cite{s_bryant1988spin}.
\par
The second order Hamiltonian is given by 
    \begin{align*}
        \begin{split}
            \hat{H}^{(2)} =& -S\sum_{j,j'} J_{jj'} \left( 2 \hat{a}_j \hat{a}_{j'}^{\dagger} - \hat{a}_{j'}^{\dagger} \hat{a}_{j'} - \hat{a}_j^{\dagger} \hat{a}_j \right) + \gamma H_0^z \sum_j \hat{a}_j^{\dagger} \hat{a}_j, \\
            & + \frac{S \gamma^2}{2} \sum_{j,j'} \frac{1}{r_{jj'}^3} \Biggl[ - \left( 1 - \frac{3 z_{jj'}^2}{r_{jj'}^2} \right) \hat{a}_j^{\dagger} \hat{a}_j - \left( 1 - \frac{3 z_{jj'}^2}{r_{jj'}^2} \right) \hat{a}_{j'}^{\dagger} \hat{a}_{j'} + \left( 1 - \frac{3}{2} \frac{r_{jj'}^+ r_{jj'}^-}{r_{jj'}^2} \right) \hat{a}_j^{\dagger} \hat{a}_{j'} \\
            & + \left( 1 - \frac{3}{2} \frac{r_{jj'}^+ r_{jj'}^-}{r_{jj'}^2} \right) \hat{a}_j \hat{a}_{j'}^{\dagger} - \frac{3}{2} \frac{ \left( r_{jj'}^- \right)^2 }{ r_{jj'}^2 } \hat{a}_j \hat{a}_{j'} - \frac{3}{2} \frac{ \left( r_{jj'}^+ \right)^2 }{ r_{jj'}^2 } \hat{a}_j^{\dagger} \hat{a}_{j'}^{\dagger} \Biggr], \\
            =& \sum_{\bf k} \left( \gamma H_0^z + 2 S [ J(0) - J({\bf k}) ] - \frac{ S \gamma^2 }{ 2 a_0^3 } \left[ D_1({\bf k}) - D_1(0) \right] \right) \hat{a}_{\bf k}^{\dagger} \hat{a}_{\bf k} - \frac{1}{2} \sum_{\bf k} \left[ \frac{ 3 S \gamma^2 }{ 2 a_0^3 } D_2({\bf k}) \hat{a}_{\bf k} \hat{a}_{ - {\bf k} } + \text{h.c.} \right]
        \end{split}
    \end{align*}
    where for small wave numbers, $J(0)-J({\bf k})=J(ka)^2$, and
    \begin{align}
        J({\bf k})=\frac{1}{N}\sum_{j,j'}J_{jj'}\exp(\ii{\bf k}\cdot{\bf r}_{jj'})=2J[\cos(k_x a_0)+\cos(k_y a_0)+\cos(k_z a_0)].
    \end{align}
Applying the dipolar sums in Appendix.(\ref{append:dipolarsum}), we can derive 
\begin{align}
    \hat{H}^{(2)}=&
    \sum_{\bf k}\left[A_{\bf k}\hat{a}_{\bf k}^{\dagger}\hat{a}_{\bf k}+\frac{1}{2}B_{\bf k}^*\hat{a}_{\bf k}\hat{a}_{\bf k}+\frac{1}{2}B_{\bf k}\hat{a}_{\bf k}^{\dagger}\hat{a}_{\bf k}^{\dagger}\right],\label{eq:secondorder}
\end{align}
where
\begin{align}
    A_{\bf k}=&\gamma H_0^z+2JS(ka_0)^2+2\pi\gamma M\left[F_{\bf k}+\sin^2(\theta_{\bf k})(1-F_{\bf k})\right]\\
    B_{\bf k}=&2\pi\gamma M\left[F_{\bf k}-\sin^2(\theta_{\bf k})(1-F_{\bf k})\right],
\end{align}
$a_0$ is the lattice spacing, $M=\gamma S/a_0^3$ is the saturation magnetization, $\theta_{\bf k}$ and $\varphi_{\bf k}$ denote the polar and the azimuthal angles between wave vectors correspondingly, the external field $H_0^z$ is directed along the $z$-axis,  $F_{\bf k}=\qty(1-e^{-\abs{\vb{k}}d})/\abs{\vb{k}}d$, and $d$ is the film thickness \cite{s_rezende2009theory}. Taking into account the film thickness explicitly is important, due to the long-range character of dipole-dipole interactions. Eq. (\ref{eq:secondorder}) can be diagonalized using the Bogoliubov transformation:
\begin{subequations}
    \begin{align}
    \hat{a}_{\mathbf{k}}& =u_{\mathbf{k}}\hat{c}_{\mathbf{k}}+v_{\mathbf{k}}\hat{c}_{-\mathbf{k}}^{\dagger}&\hat{a}_{\mathbf{k}}^{\dagger}&=u_{\mathbf{k}}\hat{c}_{\mathbf{k}}^{\dagger}+v_{\mathbf{k}}^*\hat{c}_{-\mathbf{k}}\\
    u_{\mathbf{k}}&= \sqrt{\frac{A_{\mathbf{k}}+\omega_{\mathbf{k}}}{2\omega_{\mathbf{k}}}}& v_{\mathbf{k}}&=\frac{B_{\mathbf{k}}}{|B_{\mathbf{k}}|}\sqrt{\frac{A_{\mathbf{k}}-\omega_{\mathbf{k}}}{2\omega_{\mathbf{k}}}}
\end{align}
\end{subequations}
such that $|u_{\mathbf{k}}|^2-|v_{\mathbf{k}}|^2=1$. This leads to
\begin{align}
    \hat{H}^{(2)}&= \sum_{\mathbf{k}}\omega_{\mathbf{k}}\hat{c}_{\mathbf{k}}^{\dagger}\hat{c}_{\mathbf{k}},
    \label{eq:freeHami}
\end{align}
with $\omega_{\mathbf{k}}^2=A_{\mathbf{k}}^2-|B_{\mathbf{k}}|^2$.

\section{Magnon-phonon Interactions}\label{append:mpinteract}

Additional second- and third-order interactions by phonon bath should be considered \cite{s_bozhko2017bottleneck}. One can expand the exchange coupling $J_{jj'}$ in powers of the lattice displacements \cite{s_kreisel2009microscopic, s_ruckriegel2014magnetoelastic}
\begin{align}
    J(r_{jj'})=&J\exp\left(-\kappa\frac{r_{jj'}-R_{jj'}}{R_{jj'}}\right)\\
    =&\sum_{l=0}\frac{1}{l!}({\bf u}_{jj'}\cdot\nabla)^lJ(r)\bigr\vert_{r=R_{jj'}}=J-\frac{\kappa J}{R_{jj'}}\hat{\bf x}_{jj'}\cdot{\bf u}_{jj'}+\ldots\label{eq:exchangeexpand}
\end{align}
where $r_{jj'}$ denotes the actual distance, $R_{jj'}$ denotes the distance in equilibrium and $\kappa$ is a dimensionless constant, $\hat{\bf x}$ is the unit vector connecting two sides of the cubic lattice, and ${\bf u}_{jj'}$ denotes the lattice
displacement in terms of phonon operators \cite{s_huser2016kinetic}. The exchange contribution to the magnon-phonon interactions reads ($k,k'$ mean ${\abs{\vb{k}},\abs{\vb{k}'}}$)\cite{s_gurevich1996magnetization}
\begin{align}
    \hat{H}_{\rm m,ph,ex}=\frac{1}{\sqrt{N}}\sum_{\bf k,k',\sigma}JSa_0^2\sqrt{\frac{2\hbar}{mv_{\sigma}}}kk'\sqrt{k-k'}\hat{c}_{\bf k}^{\dagger}\hat{c}_{\bf k'}\left(\hat{b}_{\bf k'-k\sigma}^{\dagger}+\hat{b}_{\bf k-k'\sigma}\right).
\end{align}
The dipole-dipole and spin-orbit contribution was investigated in Ref. \cite{s_ruckriegel2015rayleigh, s_ruckriegel2014magnetoelastic, s_streib2019magnon}. With that, the interaction Hamiltonian reads
\begin{align}
\begin{split}
    \hat{H}_{\rm m,ph,dip}=&\sum_{{\bf k},\sigma}\frac{\hat{b}_{\bf -k\sigma}+\hat{b}_{\bf k\sigma}^{\dagger}}{\sqrt{2m \omega_{\bf-k\sigma}}}(\Gamma_{\bf k\sigma}\hat{a}_{\bf k}+\Gamma_{\bf -k\sigma}^*\hat{a}_{\bf -k}^{\dagger})+\frac{1}{\sqrt{N}}\sum_{{\bf k,q},\sigma}\frac{\hat{b}_{\bf -q\sigma}+\hat{b}_{\bf q\sigma}^{\dagger}}{\sqrt{2m \omega_{\bf -q\sigma}}}{\bf e}_{\bf -q\sigma}\\
    &\cdot\left[{\bf U}_{\bf -q}\hat{a}_{\bf k}^{\dagger}\hat{a}_{\bf k+q}+\frac{1}{2}\left({\bf V}_{\bf -q}\hat{a}_{\bf -k}\hat{a}_{\bf k+q}+{\bf V}_{\bf q}^*\hat{a}_{\bf k}^{\dagger}\hat{a}_{\bf -k-q}^{\dagger}\right)\right]
\end{split}
\end{align}
where $\sigma=\{\parallel,\perp_1,\perp_2\}$ labels three acoustic phonon polarization, $\omega_{\bf k\sigma}$ are phonon frequencies, and $m$ is the effective ionic mass in a unit cell. $\{\hat{b}_{\bf k\sigma}^{\dagger},\hat{b}_{\bf k\sigma}\}$ are bosonic creation and annihilation
operators of the phonon eigen-modes with momentum ${\bf k}$ and polarization ${\bf e}_{\bf k\sigma}$. The phonon Hamiltonian gives \cite{s_ruckriegel2014magnetoelastic}
\begin{align}
    \hat{H}_{\rm ph}=\sum_{\bf k,\sigma}\omega_{\bf k\sigma}\left(\hat{b}_{\bf k\sigma}^{\dagger}\hat{b}_{\bf k\sigma}+\frac{1}{2}\right).
\end{align}
The values of the effective ionic mass gives
\begin{align}
    m=&\rho a_0^3&\rho=5.17\ {\rm g/cm^3}
\end{align}
and the longitudinal and transverse phonon velocities are defined via $\omega_{\bf k\sigma}=v_{\sigma}|{\bf k}|$ and
\begin{align}
    v_{\parallel}=&7.209\times10^5\ {\rm cm/s}&v_{\perp}=&3.843\times10^5\ {\rm cm/s}.
\end{align}
The polarization vector is defined as follows
\begin{align}
    {\bf e}_{\bf k\parallel}=&\ii({\bf e}_{z}\cos{\theta_{\bf k}}+{\bf e}_y\sin{\theta_{\bf k}})&{\bf e}_{\bf k\perp_1}=&\ii({\bf e}_{z}\sin{\theta_{\bf k}}-{\bf e}_y\cos{\theta_{\bf k}})&{\bf e}_{\bf k\perp_2}=&{\bf e}_{x},
\end{align}
and the coefficients read \cite{s_huser2016kinetic, s_ruckriegel2014magnetoelastic}
\begin{align}
    {\bf U}_{\bf q}=&\frac{\ii B_{\parallel}}{S}(q_x{\bf e}_x+q_y{\bf e}_y+q_z{\bf e}_z)\\
    {\bf V}_{\bf q}=&\frac{\ii B_{\parallel}}{S}(q_x{\bf e}_x-q_y{\bf e}_y)+\frac{ B_{\perp}}{S}(q_x{\bf e}_y+q_y{\bf e}_x)
\end{align}
with the relevant coupling coefficients
\begin{align}
    B_{\parallel}=&3.48\times10^6\times a_0^3\ {\rm erg/cm^3}&B_{\perp}=&6.96\times10^6\times a_0^3\ {\rm erg/cm^3}.
\end{align}
After the Bogoliubov transformation, the magnon-phonon interaction term gives
\begin{align}
    \begin{split}
    \hat{H}_{\rm m,ph,dip}=&\sum_{{\bf k},\sigma}\frac{1}{\sqrt{2m \omega_{\bf -k\sigma}}}[\hat{b}_{\bf -k\sigma}(\Gamma_{\bf k\sigma}v_{\bf k}+\Gamma_{\bf -k\sigma}^*u_{\bf-k})\hat{c}_{\bf -k}^{\dagger}+\hat{b}_{\bf k\sigma}^{\dagger}(\Gamma_{\bf k\sigma}u_{\bf k}+\Gamma_{\bf -k\sigma}^*v_{\bf-k}^*)\hat{c}_{\bf k}]\\
    &+\frac{1}{\sqrt{N}}\sum_{{\bf k,q},\sigma}\left[\frac{\hat{b}_{\bf k-k'\sigma}+\hat{b}_{\bf k'-k\sigma}^{\dagger}}{\sqrt{2m \omega_{\bf k-k'\sigma}}}{\bf e}_{\bf k-k'\sigma}\cdot {\boldsymbol \Gamma}_{\bf k,k'}^{c'c}\hat{c}_{\bf k}^{\dagger}\hat{c}_{\bf k'}\right.\\
    &\left.+\frac{1}{2}\left(\frac{\hat{b}_{\bf -k-k'\sigma}+\hat{b}_{\bf k'+k\sigma}^{\dagger}}{\sqrt{2m \omega_{\bf -k-k'\sigma}}}{\bf e}_{\bf k-k'\sigma}\cdot{\boldsymbol \Gamma}_{\bf k,k'}^{cc}\hat{c}_{\bf k}\hat{c}_{\bf k'}+\frac{\hat{b}_{\bf k+k'\sigma}+\hat{b}_{\bf -k-k'\sigma}^{\dagger}}{\sqrt{2m \omega_{\bf k+k'\sigma }}}{\bf e}_{\bf -q\sigma}\cdot{\boldsymbol \Gamma}_{\bf k,k'}^{c'c'}\hat{c}_{\bf k}^{\dagger}\hat{c}_{\bf k'}^{\dagger}\right)\right]
\end{split}
\end{align}
where
\begin{align}
    {\boldsymbol\Gamma}_{\bf k,k'}^{c'c}=&(u_{\bf k}u_{\bf k'}+v_{\bf k}v_{\bf k'}^*){\bf U}_{\bf k-k'}+v_{\bf k}u_{\bf k'}{\bf V}_{\bf k'-k}+u_{\bf k}v_{\bf k'}^*{\bf V}_{\bf k'-k}^*\\
    {\boldsymbol\Gamma}_{\bf k,k'}^{cc}=&(v_{\bf k}^*u_{\bf k'}+u_{\bf k}v_{\bf k'}^*){\bf U}_{\bf -k-k'}+u_{\bf k}u_{\bf k'}{\bf V}_{\bf k'+k}+v_{\bf k}^*v_{\bf k'}^*{\bf V}_{\bf k'+k}^*\\
    {\boldsymbol\Gamma}_{\bf k,k'}^{c'c'}=&(u_{\bf k}v_{\bf k'}+u_{\bf k'}v_{\bf k}){\bf U}_{\bf k+k'}+u_{\bf k}u_{\bf k'}{\bf V}_{\bf -k'-k}^*+v_{\bf k}v_{\bf k'}{\bf V}_{\bf -k'-k}.
\end{align}
Here, if we do not consider higher-order magnon-magnon interactions (higher than the second order) but only combine the exchange and dipolar contributions to magnon-phonon interactions, the interaction Hamiltonian in the interaction picture reads
\begin{align}
\begin{split}
    \hat{H}_{\rm m,ph}=&\sum_{{\bf k},\sigma}V_{\sigma}({\bf k})\hat{b}_{\bf k\sigma}\hat{c}_{\bf k}^{\dagger}e^{\ii(\omega_{\bf k}-\omega_{\bf k\sigma})t}+\text{h.c.}\\
    &+\frac{1}{\sqrt{N}}\sum_{\bf k,k',\sigma}V_{\sigma}({\bf k,k'})\hat{c}_{\bf k}^{\dagger}\hat{c}_{\bf k'}\hat{b}_{\bf k-k'\sigma}e^{\ii(\omega_{\bf k}-\omega_{\bf k'}-\omega_{\bf k-k'\sigma})t}+\text{h.c.}
\end{split}\label{eq:s_magnonphononinterac}
\end{align}
where
\begin{align}
    V_{\sigma}({\bf k})=&\frac{\Gamma_{\bf -k\sigma}v_{\bf -k}+\Gamma_{\bf k\sigma}^*u_{\bf k}}{\sqrt{2m \omega_{\bf k\sigma}}}\\
    V_{\sigma}({\bf k,k'})=&JSa_0^2\sqrt{\frac{2\hbar}{mv_{\sigma}}}kk'\sqrt{k-k'}+\frac{{\bf e}_{\bf k-k'\sigma}\cdot {\boldsymbol \Gamma}_{\bf k,k'}^{c'c}}{\sqrt{2m \omega_{\bf k-k'\sigma}}}.
\end{align}

\section{Derivation of the Fr\"{o}hlich rate equation using the theory of open systems}
In the interaction picture, the total interaction Hamiltonian is of general form including the pumping, linear, and nonlinear terms ($C_k\equiv V_{\sigma}({\bf k})$ and ${\cal C}_{kj}\equiv V_{\sigma}({\bf k,k'})$),
\begin{align}
\begin{split}
     \hat{V}_{\text{I}}(t)=&\sum_{k} \left[g_k\hat{a}^{\dagger}_ke^{\ii(\omega_k-\omega_{\rm p})t}+\text{h.c.}\right]+\sum_{l,k}\left[C_k\hat{a}_k^{\dagger}\hat{b}_le^{\ii(\omega_k-\Omega_l) t}+\text{h.c.}\right]+\sum_{l,k}\sum_{j\neq k}\left[\mathcal{C}_{kj}\hat{a}_k^{\dagger}\hat{a}_j\hat{b}_le^{\ii(\omega_k-\omega_j-\Omega_l)t}+\text{h.c.}\right]
\end{split}\label{eq:interactionHamiapproxpump}
\end{align}
where the system oscillators of operators, say $\hat{a}_k^{\dagger}$, correspond to frequencies $\omega_k$ the bath frequencies are $\hat{b}_l^{\dagger}\Leftrightarrow\Omega_l$, the time-dependent pumping has frequency $\omega_{\rm p}$ that can be merged into coefficient $g_k$.
The LME can then be derived
\begin{align}
    \begin{split}
        \dv{t}\hat{\rho}_{\mathrm{S}}(t)=&\sum_{k}s_k\left(\hat{a}_k^{\dagger}\hat{\rho}_{\text{S}}(t)\hat{a}_k-\frac{1}{2}\left\{\hat{a}_k\hat{a}_k^{\dagger},\hat{\rho}_{\text{S}}(t)\right\}\right)+\sum_{k}s_k\left(\hat{a}_k\hat{\rho}_{\text{S}}(t)\hat{a}_k^{\dagger}-\frac{1}{2}\left\{\hat{a}_k^{\dagger}\hat{a}_k,\hat{\rho}_{\text{S}}(t)\right\}\right)\\
        &+\sum_{k,l}\gamma_{\text{lin}}\bar{n}_{\text{B}}(\Omega_l)\left(\hat{a}_k^{\dagger}\hat{\rho}_{\text{S}}(t)\hat{a}_k-\frac{1}{2}\left\{\hat{a}_k\hat{a}_k^{\dagger},\hat{\rho}_{\text{S}}(t)\right\}\right)\\
        &+\sum_{k,l}\gamma_{\text{lin}}(\bar{n}_{\text{B}}(\Omega_l)+1)\left(\hat{a}_k\hat{\rho}_{\text{S}}(t)\hat{a}_k^{\dagger}-\frac{1}{2}\left\{\hat{a}_k^{\dagger}\hat{a}_k,\hat{\rho}_{\text{S}}(t)\right\}\right)\\
        &+\sum_{k,l}\sum_{j\neq k}\gamma_{\text{non}}\bar{n}_{\text{B}}(\Omega_l)\left(\hat{a}_k^{\dagger}\hat{a}_j\hat{\rho}_{\text{S}}(t)\hat{a}_k\hat{a}_j^{\dagger}-\frac{1}{2}\left\{\hat{a}_k\hat{a}_j^{\dagger}\hat{a}_k^{\dagger}\hat{a}_j,\hat{\rho}_{\text{S}}(t)\right\}\right)\\
    &+\sum_{k,l}\sum_{j\neq k}\gamma_{\text{non}}(\bar{n}_{\text{B}}(\Omega_l)+1)\left(\hat{a}_k\hat{a}_j^{\dagger}\hat{\rho}_{\text{S}}(t)\hat{a}_k^{\dagger}\hat{a}_j-\frac{1}{2}\left\{\hat{a}_k^{\dagger}\hat{a}_j\hat{a}_k\hat{a}_j^{\dagger},\hat{\rho}_{\text{S}}(t)\right\}\right),
    \end{split}\label{eq:qmsFrohHami}
\end{align}
where
\begin{align}
    \gamma_{\text{lin}}=&\abs{C_k}^2\int_{-\infty}^{\infty}\mathrm{d}se^{\ii(\omega_k-\Omega_l)s}=2\pi\abs{C_k}^2\delta(\omega_k-\Omega_l)&
    \gamma_{\text{non}}=&\abs{\mathcal{C}_{kj}}^2\int_{-\infty}^{\infty}\mathrm{d}se^{\ii(\Delta_{kj}-\Omega_l)s}=2\pi \abs{\mathcal{C}_{kj}}^2\delta(\Delta_{kj}-\Omega_l).
\end{align}
The integral in the coefficients $\gamma_{\text{lin}}$ and $\gamma_{\text{non}}$ can be carrier out to be Dirac delta functions. This refers to the Fermi golden rule, as the Markov approximation extends the integral limits to infinity.
\par 
The corresponding rate equation results in a constant pumping term
\begin{align}
    \begin{split}
        \dv{t}\expval{\hat{n}_k}=&s_k\tr_{\rm S}\qty{\hat{n}_k\hat{a}_k^{\dagger}\hat{\rho}_{\rm S}\hat{a}_k-\hat{n}_k\frac{1}{2}\qty{\hat{a}_k\hat{a}_k^{\dagger},\hat{\rho}_{\rm S}}+\hat{n}_k\hat{a}_k\hat{\rho}_{\rm S}\hat{a}_k^{\dagger}-\hat{n}_k\frac{1}{2}\qty{\hat{a}_k^{\dagger}\hat{a}_k,\hat{\rho}_{\rm S}}}\\
        =& s_k\tr_{\rm S}\qty{\qty[\hat{a}_k\hat{n}_k\hat{a}_k^{\dagger}-\hat{n}_k\hat{a}_k\hat{a}_k^{\dagger}+\hat{a}_k^{\dagger}\hat{n}_k\hat{a}_k-\hat{n}_k\hat{a}_k^{\dagger}\hat{a}_k]\hat{\rho}_{\rm S}}\\
        =& s_k\expval{\hat{a}_k\hat{n}_k\hat{a}_k^{\dagger}-\hat{n}_k\hat{a}_k\hat{a}_k^{\dagger}+\hat{a}_k^{\dagger}\hat{n}_k\hat{a}_k-\hat{n}_k\hat{a}_k^{\dagger}\hat{a}_k}\\
        =&s_k\expval{[\hat{a}_k,\hat{n}_k]\hat{a}_k^{\dagger}+[\hat{a}_k^{\dagger},\hat{n}_k]\hat{a}_k}\\
        =&s_k\expval{\hat{a}_k\hat{a}_k^{\dagger}-\hat{a}_k^{\dagger}\hat{a}_k}\\
        =&s_k\expval{[\hat{a}_k,\hat{a}_k^{\dagger}]}\\
        =&s_k,
    \end{split}\label{eq:pumpderive}
\end{align}
where we have applied the cyclic property of the trace, and the coefficient is provided by the Fermi Golden rule (this can be replaced by an integral of density of magnon states over frequencies $\omega_{\bf k}$, but we adopt the delta function for simplicity):
\begin{align}
    s_k&=\abs{g_k}^2\int_{-\infty}^{\infty}\mathrm{d}se^{\ii\omega_{k}s}=2\pi\abs{g_k}^2\delta(\omega_k).
\end{align}
The derivation of the rate equation for the linear transitions can be analogously performed:
\begin{align}
    \begin{split}
        \dv{t}\expval{\hat{n}_k}=&\gamma_{\rm lin}\bar{n}_{\rm B}\expval{\hat{a}_k\hat{n}_k\hat{a}_k^{\dagger}-\hat{n}_k\hat{a}_k\hat{a}_k^{\dagger}+e^{\beta\omega_k}\qty(\hat{a}_k^{\dagger}\hat{n}_k\hat{a}_k-\hat{n}_k\hat{a}_k^{\dagger}\hat{a}_k)}\\
        =&\gamma_{\rm lin}\bar{n}_{\rm B}\expval{[\hat{a}_k,\hat{n}_k]\hat{a}_k^{\dagger}+e^{\beta\omega_k}[\hat{a}_k^{\dagger},\hat{n}_k]\hat{a}_k}\\
        =&\gamma_{\rm lin}\bar{n}_{\rm B}\expval{\hat{a}_k\hat{a}_k^{\dagger}-e^{\beta\omega_k}\hat{a}_k^{\dagger}\hat{a}_k}\\
        =&\varphi_k(\beta)\qty(\expval{\hat{n}_k+\hat{1}}-\expval{\hat{n}_k}e^{\beta\omega_k})\\
        =&\varphi_k(\beta)\qty(\expval{\hat{n}_k}+1-\expval{\hat{n}_k}e^{\beta\omega_k}),
    \end{split}\label{eq:linderive}
\end{align}
where, similarly, $\varphi_k(\beta)\equiv \gamma_{\rm lin}\bar{n}_{\rm B}(\Omega_l) = 2\pi \abs{C_{k}}^2\bar{n}_{\rm B}(\Omega_l)\delta(\omega_k-\Omega_l)=2\pi \abs{C_{k}}^2\bar{n}_{\rm B}(\omega_k)$.
\par 
One might notice the difference between the derivations of pumping term and linear transition term. The difference lies in the bath correlation functions. The coupling to pumping source reduces from operators $\{\hat{R},\hat{R}^{\dagger}\}$ to complex amplitudes which has been merged in $g_{\bf k}$ in Eq. (\ref{eq:interactionHamiapproxpump}), while the coupling to bath for linear transitions preserve the bath operators $\{\hat{b}_l,\hat{b}_l^{\dagger}\}$. This is because we assume that pumping source is different than the heat bath and is some coherent electromagnetic field, which has a very large mean photon number $\abs{\alpha}^2$, such that 
\begin{align}
\begin{split}
    &\expval*{\hat{R}^{\dagger}\hat{R}}\approx\expval*{\hat{R}\hat{R}^{\dagger}}\\
    \Leftrightarrow\ \ &\bar{n}_{\rm R}(\omega_{\rm p})\approx\bar{n}_{\rm R}(\omega_{\rm p})+1\\
    \Leftrightarrow\ \ &\abs{\alpha}^2
     \approx\abs{\alpha}^2+1.\\
\end{split}
\end{align}
Note that $\bar{n}_{\rm R}(\omega_{\rm p})$ is the Planck distribution for the resonator.
Technically, in derivations in Eq. (\ref{eq:pumpderive} and \ref{eq:linderive}), there is no exponential factor $e^{\beta\omega_{\bf k}}$ coming out from 
\begin{align}
    \delta(\omega_k-\Omega_l)[\bar{n}_{\rm B}(\Omega_l)+1]=\frac{e^{\beta \omega_k}}{e^{\beta \omega_k}-1}=\bar{n}_{\rm B}(\omega_k)e^{\beta \omega_k},
\end{align}
when we trace out the bath degrees of freedom.
\par 
Recall that Preto's and other relevant attempts \cite{s_preto2017semi, s_wu1977bose, s_wu1978bose, s_wang2022full} derived the pumping term $s_k$ in the semi-classical model through a linear transition with the pumping source at infinite temperature, 
\begin{align}
    s_k\equiv\lim_{\beta_{\rm p}\to0}s_k\left(\expval{n_k}+1-\expval{n_k}e^{\beta_{\rm p}\omega_k}\right),
\end{align}
where $\beta_{\rm p}$ is the temperature of pumping source. For the nonlinear transitions, we have
\begin{align}
    \begin{split}
        \dv{t}\expval{\hat{n}_k}=&\sum_{j\neq k}\gamma_{\rm non}\bar{n}_{\rm B}\expval{(\hat{a}_k\hat{n}_k\hat{a}_k^{\dagger})\otimes(\hat{a}_j^{\dagger}\hat{a}_j)-(\hat{n}_k\hat{a}_k\hat{a}_k^{\dagger})\otimes(\hat{a}_j^{\dagger}\hat{a}_j)}\\
        &+\sum_{j\neq k}\gamma_{\rm non}\bar{n}_{\rm B}e^{\beta(\omega_k-\omega_j)}\expval{(\hat{a}_k^{\dagger}\hat{n}_k\hat{a}_k)\otimes(\hat{a}_j\hat{a}_j^{\dagger})-(\hat{n}_k\hat{a}_k^{\dagger}\hat{a}_k)\otimes(\hat{a}_j\hat{a}_j^{\dagger})}\\
        =&\sum_{j\neq k}\gamma_{\rm non}\bar{n}_{\rm B}\qty[\expval{([\hat{a}_k,\hat{n}_k]\hat{a}_k^{\dagger})\otimes(\hat{a}_j^{\dagger}\hat{a}_j)}+e^{\beta(\omega_k-\omega_j)}\expval{([\hat{a}_k^{\dagger},\hat{n}_k]\hat{a}_k)\otimes(\hat{a}_j\hat{a}_j^{\dagger})}]\\
        =&\sum_{j\neq k}\gamma_{\rm non}\bar{n}_{\rm B}\qty[\expval{(\hat{a}_k\hat{a}_k^{\dagger})\otimes(\hat{a}_j^{\dagger}\hat{a}_j)}+e^{\beta(\omega_k-\omega_j)}\expval{(-\hat{a}_k^{\dagger}\hat{a}_k)\otimes(\hat{a}_j\hat{a}_j^{\dagger})}]\\
        =& \sum_{j\neq k}\gamma_{\rm non}\bar{n}_{\rm B}[-e^{\beta(\omega_k-\omega_j)}\expval{\hat{n}_k(\hat{n}_j+\hat{1})}+\expval{(\hat{n}_k+\hat{1})\hat{n}_j}]\\
        =&\sum_{j\neq k}\Lambda_{kj}(\beta)[\expval{(\hat{n}_k+\hat{1})\hat{n}_j}-\expval{\hat{n}_k(\hat{n}_j+\hat{1})}e^{\beta(\omega_k-\omega_j)}]
    \end{split}\label{eq:nonlinderive}
\end{align}
where $\Lambda_{kj}(\beta)\equiv\gamma_{\rm non}\bar{n}_{\rm B}(\Omega_l) =2\pi \abs{\mathcal{C}_{kj}}^2\bar{n}_{\rm B}(\Omega_l)\delta(\omega_k-\omega_j-\Omega_l) = 2\pi \abs{\mathcal{C}_{kj}}^2\bar{n}_{\rm B}(\omega_k-\omega_j)$. Combining Eq. (\ref{eq:pumpderive}), (\ref{eq:linderive}), and (\ref{eq:nonlinderive}), we can reach a quantum Fr\"{o}hlich rate equation,
\begin{equation}
    \begin{split}
        \dv{t}\expval{\hat{n}_k}=& s_k+\varphi_k(\beta)\left(\expval{\hat{n}_k}+1-\expval{\hat{n}_k}e^{\beta\omega_k}\right)+\sum_{j\neq k}\Lambda_{kj}(\beta)\left[(\expval{\hat{n}_k}+1)\expval{\hat{n}_j}-\expval{\hat{n}_k}(\expval{\hat{n}_j}+1)e^{\beta(\omega_k-\omega_j)}\right],
    \end{split}\label{eq:quantumFroh}
\end{equation}
where the factorized-state approximation $\expval{\hat{n}_k\hat{n}_j}\approx\expval{\hat{n}_k}\expval{\hat{n}_j}$ with $k\neq j$ has been applied. 

\nocite{*}

\providecommand{\noopsort}[1]{}\providecommand{\singleletter}[1]{#1}%

\end{document}